\documentclass[12pt,a4paper]{article}
\usepackage{amsmath}
\usepackage{graphicx}
\begin{document}
\title{
Ward identities \\ 
for particle-particle and particle-hole pairs
\footnote{In v2 eqs.~(24) and (25) have been added and footnote-2 
has been corrected.}
}

\author{
O. Narikiyo
\footnote{
Department of Physics, 
Kyushu University, 
Fukuoka 812-8581, 
Japan}
}

\date{
(Oct. 5, 2011)
}

\maketitle
\begin{abstract}
Ward identities for charge-density and spin-density fluctuations 
are discussed in comparison with 
those for superconducting fluctuations. 
\end{abstract}

\vskip 30pt 

\section{Introduction}

Recently I have reported the discussion of Ward identities 
for superconducting fluctuations~\cite{Nar1,Nar2}. 
In this note 
I will extend it to the case of charge-density and spin-density fluctuations. 
While the superconducting fluctuation is described 
by particle-particle pair propagator, 
the charge-density and spin-density fluctuations are described 
by particle-hole pair propagator. 

For the sake of clarity 
I will only discuss the case of zero temperature. 
At finite temperature 
we can use the same Ward identity with thermal frequency~\cite{Nar1}. 
In addition I will only discuss the local pairs for simplicity. 
The discussion of nonlocal pairs has been already given in ref.~\cite{Nar2}. 

Following description is based on refs.~\cite{Nar1} and \cite{Nar2}. 

\section{Ward identity for electric current vertex}

Under the charge-conservation law 
\begin{equation}
\sum_{\mu=0}^3 {\partial \over \partial z_\mu} j_\mu^e(z) = 0, \label{cc-law}
\end{equation}
the three-point function $M_\mu^e(x,y,z)$ shown in Fig.~1 
\begin{equation}
M_\mu^e(x,y,z) = \big\langle T 
\{ j_\mu^e(z) A(x) A^\dag(y) \} \big\rangle,
\end{equation}
satisfies the relation 
\begin{align}
\sum_{\mu=0}^3 {\partial \over \partial z_\mu} M_\mu^e(x,y,z) = 
& \Big\langle T 
\Big\{ \big[ j_0^e(z), A(x) \big] A^\dag(y) \Big\} \Big\rangle 
\delta(z_0-x_0) \nonumber \\ 
+ 
& \Big\langle T 
\Big\{ A(x) \big[ j_0^e(z), A^\dag(y) \big] \Big\} \Big\rangle 
\delta(z_0-y_0), \label{divM-e-coordinate} 
\end{align}
where 
\begin{equation}
j_0^e(z) = 
e \psi_\uparrow^\dag(z) \psi_\uparrow(z) + 
e \psi_\downarrow^\dag(z) \psi_\downarrow(z), 
\end{equation}
and 
$A(x)$ and $A^\dag(y)$ 
represent the annihilation and creation operators 
of particle-particle pairs 
\begin{equation}
\Psi^\dag(x) \equiv \psi_\uparrow^\dag(x) \psi_\downarrow^\dag(x), 
\ \ \ \ \ 
\Psi(x) \equiv \psi_\downarrow(x) \psi_\uparrow(x), 
\end{equation}
or particle-hole pairs 
\begin{equation}
\rho_\uparrow(x) \equiv \psi_\uparrow^\dag(x) \psi_\uparrow(x), 
\ \ \ \ \ 
\rho_\downarrow(x) \equiv \psi_\downarrow^\dag(x) \psi_\downarrow(x), 
\end{equation}
\begin{equation}
\sigma_+(x) \equiv \psi_\uparrow^\dag(x) \psi_\downarrow(x), 
\ \ \ \ \ 
\sigma_-(x) \equiv \psi_\downarrow^\dag(x) \psi_\uparrow(x). 
\end{equation}
\vskip 8mm
\begin{figure}[htbp]
\begin{center}
\includegraphics[width=3.6cm]{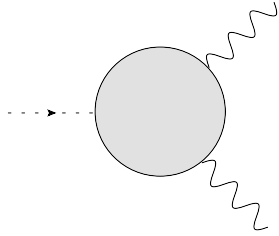}
\vskip 4mm
\caption{Feynman diagram for three-point function $M_\mu^e$: 
The shaded circle represents the coupling of 
the external electromagnetic field (broken line) 
and the particle-particle or particle-hole fluctuations (wavy lines). 
All the Feynman diagrams in this note 
are drawn by JaxoDraw~\cite{BT}. }
\label{fig:Fig1}
\end{center}
\end{figure}
Using the commutation relations for particle-particle pairs 
\begin{equation}
[j_0^e(z), A(x)] \delta(z_0-x_0) = -2e A(x) \delta^4(z-x), 
\end{equation}
\begin{equation}
[j_0^e(z), A^\dag(y)] \delta(z_0-y_0) = 2e A^\dag(y) \delta^4(z-y), 
\end{equation}
we have obtained the Ward identity 
for particle-particle pairs~\cite{Nar1,Nar2}. 
On the other hand, the commutation relations for particle-hole pairs 
\begin{equation}
[j_0^e(z), A(x)] \delta(z_0-x_0) = 0, 
\end{equation}
\begin{equation}
[j_0^e(z), A^\dag(y)] \delta(z_0-y_0) = 0, 
\end{equation}
with 
$ j_0^e(z) = e \big( \rho_\uparrow(z) + \rho_\downarrow(z) \big) $ 
lead the Ward identity for the electric current vertex 
of particle-hole pairs 
\begin{equation}
\sum_{\mu=0}^3 {\partial \over \partial z_\mu} M_\mu^e(x,y,z) = 0. 
\end{equation}
It should be noted here 
that the commutation relation picks up 
the charge\footnote{The electric charge ${\cal Q}$ is given as 
${\cal Q} = \int d{\vec z}\, \, j_0^e(z)$. 
Here we consider the charge carried by local pairs. } 
carried by the pair~\cite{IZ}. 
Particle-hole pairs are charge-neutral 
and carry no charge. 
Namely charge- and spin-density fluctuations 
do not couple to electromagnetic field. 

The relation, eq.~(\ref{divM-e-coordinate}), 
is Fourier transformed into 
\begin{align}
\sum_{\mu=0}^3 i k_\mu M_\mu^e(q,q-k) = 
\int d & (x_0-y_0) e^{-iq_0(x_0-y_0)} \int d (z_0-x_0) e^{-ik_0(z_0-x_0)} 
\nonumber \\ 
\times \Bigl(
& \Big\langle T 
\Big\{ \big[ j_{\vec k}^e(x_0), A_{{\vec q}-{\vec k}}(x_0) \big] 
A_{\vec q}^\dag(y_0) \Big\} \Big\rangle \delta(z_0-x_0) \nonumber \\ 
+ 
& \Big\langle T 
\Big\{ A_{{\vec q}-{\vec k}}(x_0) 
\big[ j_{\vec k}^e(y_0), A_{\vec q}^\dag(y_0) \big] \Big\} \Big\rangle 
\delta(z_0-y_0) \Bigr). \label{divM-e-momentum} 
\end{align}
Using the commutation relations for particle-particle pairs 
\begin{equation}
\big[ j_{\vec k}^e, A_{{\vec q}-{\vec k}} \big] 
= - 2 e A_{\vec q}, 
\ \ \ \ \ 
\big[ j_{\vec k}^e, A_{\vec q}^\dag \big] 
= 2 e A_{{\vec q}-{\vec k}}^\dag, 
\end{equation}
we have obtained the Ward identity 
for particle-particle pairs~\cite{Nar1,Nar2}. 
On the other hand, the commutation relations for particle-hole pairs 
\begin{equation}
\big[ j_{\vec k}^e, A_{{\vec q}-{\vec k}} \big] 
= 0, 
\ \ \ \ \ 
\big[ j_{\vec k}^e, A_{\vec q}^\dag \big] 
= 0, 
\end{equation}
lead the Ward identity for the electric current vertex 
of particle-hole pairs 
\begin{equation}
\sum_{\mu=0}^3 k_\mu M_\mu^e(q,q-k) = 0. \label{WI-ph}
\end{equation}

As has been discussed in ref.~\cite{Nar1} 
the current vertex for dc conductivity 
is obtained from the Ward identity 
in the limit of vanishing external momentum $ k \rightarrow 0 $. 
Thus the Ward identity, eq.~(\ref{WI-ph}), means 
that the current vertex of particle-hole pairs 
has no contribution to dc conductivity. 
Such a conclusion is natural, 
since particle-hole pairs are charge-neutral and carry no charge. 
A perturbational example of \lq\lq no contribution" 
is shown in Fig.~2. 
\vskip 14mm
\begin{figure}[htbp]
\begin{center}
\includegraphics[width=13cm]{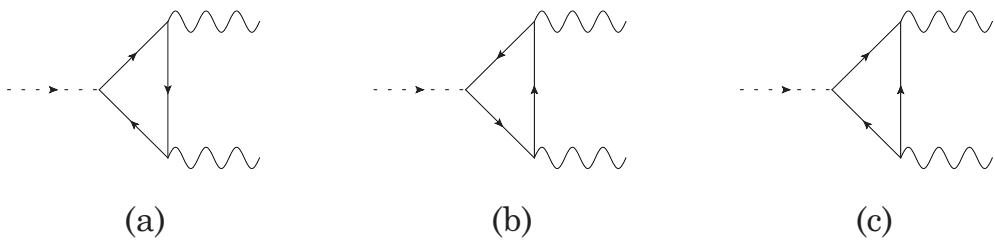}
\caption{Feynman diagrams for Aslamazov-Larkin processes: 
Wavy lines represent particle-hole fluctuation in (a) and (b) and 
particle-particle fluctuation in (c). 
Solid lines represents electron propagator. 
The contributions of (a) and (b) cancel and 
the resulting electric current vertex for particle-hole pairs vanishes 
in the case of dc conductivity. 
Such a cancelation has been discussed in ref.~\cite{Tak1} 
for spin-density fluctuation and in ref.~\cite{Tak2} 
for charge-density fluctuation. 
In the case of particle-particle pairs 
there is no canceling partner 
when three lines coupling to the triangle are fixed. 
Then superconducting fluctuation couples 
to external electromagnetic field.
}
\label{fig:Fig2}
\end{center}
\end{figure}

\newpage
\section{Ward identity for more complex vertex}

The Ward identity discussed in the previous section 
is a special case of the multi-point function $M_\mu^{(n,\, p)}$ 
\begin{equation}
M_\mu^{(n,\, p)} = 
\Big\langle T 
\Big\{ j_\mu^e(z) 
\prod_{i=1}^n \psi(x_i)\psi^\dag(y_i) 
\prod_{j=1}^p A_j(z_j) \Big\} 
\Big\rangle, 
\end{equation}
shown in Fig.~3. 
\vskip 12mm
\begin{figure}[htbp]
\begin{center}
\includegraphics[width=10cm]{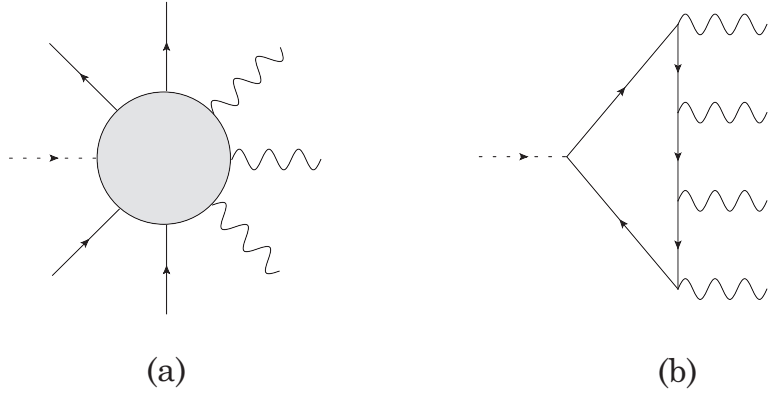}
\caption{Feynman diagrams for multi-point function $M_\mu^{(n,\, p)}$: 
(a) A general expression for $n=2$ and $p=3$. 
(b) A perturbational expression for $n=0$ and $p=4$. 
Perturbationally 
\lq\lq no contribution" of this type has been discussed 
in ref.~\cite{Tak2}. 
}
\label{fig:Fig3}
\end{center}
\end{figure}
The following derivation of the Ward identity 
is given in ref.~\cite{IZ} for QED. 
Here the spin-index of electron operator is dropped 
and the distinction between creation and annihilation operators of pairs 
is neglected. 
The derivative of the time-ordering operator $T$ leads the identity 
\begin{align}
& {\partial \over \partial t} T 
\Big\{ {\hat O}(t) {\hat O}_1(t_1) {\hat O}_2(t_2) \cdot\cdot\cdot 
{\hat O}_n(t_n) \Big\} = 
T \Big\{ {\partial {\hat O}(t) \over \partial t} 
{\hat O}_1(t_1) {\hat O}_2(t_2) \cdot\cdot\cdot {\hat O}_n(t_n) \Big\} 
\nonumber \\ 
& + \sum_{i=1}^n T 
\Big\{ {\hat O}_1(t_1) \cdot\cdot\cdot {\hat O}_{i-1}(t_{i-1}) 
\big[ {\hat O}(t), {\hat O}_i(t_i) \big] 
{\hat O}_{i+1}(t_{i+1}) \cdot\cdot\cdot {\hat O}_n(t_n) \Big\} 
\delta(t-t_i), \label{T-multi} 
\end{align}
where 
${\hat O}$ and ${\hat O}_i$ ($i=1,2,\cdot\cdot\cdot,n$) 
represent arbitrary operators. 
Making use of this identity the divergence of $M_\mu^{(n,\, p)}$ 
is shown to satisfy the relation\footnote{
Here we have introduced the notation 
\begin{align}
\big\lceil \psi(x)\psi^\dag(y) \big\rfloor & \equiv 
\big[ j_0^e(z), \psi(x) \big] \psi^\dag(y) \delta(z_0-x_0) 
+ 
\psi(x) \big[ j_0^e(z), \psi^\dag(y) \big] \delta(z_0-y_0), 
\nonumber \\ 
\big\lceil A_j(z') \big\rfloor & \equiv 
\big[ j_0^e(z), A_j(z') \big] \delta(z_0-z'_0). \nonumber
\end{align}
} 
\begin{align}
& \sum_{\mu=0}^3 {\partial \over \partial z_\mu} M_\mu^{(n,\, p)} 
\nonumber \\ 
= & \sum_{i=1}^n \Big\langle T 
\Big\{ 
\psi(x_1)\psi^\dag(y_1) \cdot\cdot\cdot \psi(x_{i-1})\psi^\dag(y_{i-1}) 
\nonumber \\ 
& \ \ \ \ \ \ \ \ \ \ \ \ \ \ \ \ \ \ \ \ \ \times
\big\lceil \psi(x_i)\psi^\dag(y_i) \big\rfloor 
\psi(x_{i+1})\psi^\dag(y_{i+1}) \cdot\cdot\cdot \psi(x_n)\psi^\dag(y_n) 
\prod_{j=1}^p A_j(z_j) 
\Big\} 
\Big\rangle 
\nonumber \\ 
+ & 
\sum_{j=1}^p \Big\langle T 
\Big\{ 
\prod_{i=1}^n \psi(x_i)\psi^\dag(y_i) 
A_1(z_1) \cdot\cdot\cdot A_{j-1}(z_{j-1}) 
\big\lceil A_j(z_j) \big\rfloor 
A_{j+1}(z_{j+1}) \cdot\cdot\cdot A_p(z_p) 
\Big\} 
\Big\rangle, \label{div-Lambda-np} 
\end{align}
under the charge-conservation law. 
The commutation relations for electrons result in 
\begin{equation}
\big[ j_0^e(z), \psi(x) \big] \delta(z_0-x_{0}) 
= - e \psi(x) \delta^4(z-x), 
\end{equation}
\begin{equation}
\big[ j_0^e(z), \psi^\dag(y) \big] \delta(z_0-y_{0}) 
= e \psi^\dag(y) \delta^4(z-y). 
\end{equation}
If we consider only particle-hole pairs, 
then the commutation relation for them results in
\begin{equation}
\big[ j_0^e(z), A_j(z') \big] \delta(z_0-z'_{0})
= 0, 
\end{equation}
where 
$A_j$ is a linear combination 
of $\{ \rho_\uparrow, \rho_\downarrow, \sigma_+, \sigma_- \}$. 
Using these commutation relations we obtain 
the Ward identity 
\begin{equation}
\sum_{\mu=0}^3 {\partial \over \partial z_\mu} M_\mu^{(n,\, p)} = e 
\Big\langle T 
\Big\{ 
\prod_{i=1}^n \psi(x_i)\psi^\dag(y_i) 
\prod_{j=1}^p A_j(z_j) \Big\} 
\Big\rangle 
\sum_{i=1}^n \Big( \delta^4(z-y_i) - \delta^4(z-x_i) \Big). 
\end{equation}
This Ward identity is identical to that in QED. 

An application of this Ward identity 
is the proof of \lq\lq no contribution" 
of multi-fluctuations of particle-hole pairs 
shown in Fig. 3. 

If we consider 
the multi-point function $M_\mu^{(n,\, m)}$ 
\begin{equation}
M_\mu^{(n,\, m)} = 
\Big\langle T 
\Big\{ j_\mu^e(z) 
\prod_{i=1}^n \psi(x_i)\psi^\dag(y_i) 
\prod_{j=1}^m A(x_j) A^\dag(y_j) \Big\} 
\Big\rangle, 
\end{equation}
where 
$A(x_j)$ and $A^\dag(y_j)$ 
are the annihilation and creation operators 
of particle-particle pairs, 
we obtain the Ward identity 
\begin{align}
\sum_{\mu=0}^3 {\partial \over \partial z_\mu} M_\mu^{(n,\, m)} 
& = e 
\Big\langle T 
\Big\{ 
\prod_{i=1}^n \psi(x_i)\psi^\dag(y_i) 
\prod_{j=1}^m A(x_j) A^\dag(y_j) \Big\} 
\Big\rangle 
\sum_{i=1}^n \Big( \delta^4(z-y_i) - \delta^4(z-x_i) \Big) 
\nonumber \\ 
& + 2e 
\Big\langle T 
\Big\{ 
\prod_{i=1}^n \psi(x_i)\psi^\dag(y_i) 
\prod_{j=1}^m A(x_j) A^\dag(y_j) \Big\} 
\Big\rangle 
\sum_{j=1}^m \Big( \delta^4(z-y_j) - \delta^4(z-x_j) \Big). 
\end{align}

\section{Ward identity for heat current vertex}

Under the energy-conservation law 
\begin{equation}
\sum_{\mu=0}^3 {\partial \over \partial z_\mu} j_\mu^Q(z) = 0, 
\end{equation}
the three-point function $M_\mu^Q(x,y,z)$ 
\begin{equation}
M_\mu^Q(x,y,z) = \big\langle T 
\{ j_\mu^Q(z) A(x) A^\dag(y) \} \big\rangle,
\end{equation}
satisfies the relation 
\begin{align}
\sum_{\mu=0}^3 {\partial \over \partial z_\mu} M_\mu^Q(x,y,z) = 
& \Big\langle T 
\Big\{ \big[ j_0^Q(z), A(x) \big] A^\dag(y) \Big\} \Big\rangle 
\delta(z_0-x_0) \nonumber \\ 
+ 
& \Big\langle T 
\Big\{ A(x) \big[ j_0^Q(z), A^\dag(y) \big] \Big\} \Big\rangle 
\delta(z_0-y_0). \label{divM-Q-coordinate} 
\end{align}
If the interaction among electrons is local, then using\footnote{
Here $j_0^Q(z)$ is the energy density and the Hamiltonian $H$ is 
given as $ H = \int d{\vec z}\, \, j_0^Q(z) $. }
\begin{equation}
\big[ j_0^Q(x), A(x) \big] = \big[ H, A(x) \big], 
\ \ \ \ \ 
\big[ j_0^Q(y), A^\dag(y) \big] = \big[ H, A^\dag(y) \big], 
\end{equation}
we easily obtain the Ward identity for heat current vertex 
as discussed in ref.~\cite{Nar1}. 
Even if the interaction is nonlocal, 
we can obtain the same Ward identity employing the Fourier transform 
\begin{align}
\sum_{\mu=0}^3 i k_\mu M_\mu^e(q,q-k) = 
\int d & (x_0-y_0) e^{-iq_0(x_0-y_0)} \int d (z_0-x_0) e^{-ik_0(z_0-x_0)} 
\nonumber \\ 
\times \Bigl(
& \Big\langle T 
\Big\{ \big[ j_{\vec k}^Q(x_0), A_{{\vec q}-{\vec k}}(x_0) \big] 
A_{\vec q}^\dag(y_0) \Big\} \Big\rangle \delta(z_0-x_0) \nonumber \\ 
+ 
& \Big\langle T 
\Big\{ A_{{\vec q}-{\vec k}}(x_0) 
\big[ j_{\vec k}^Q(y_0), A_{\vec q}^\dag(y_0) \big] \Big\} \Big\rangle 
\delta(z_0-y_0) \Bigr). \label{divM-Q-momentum} 
\end{align}
Here we can make the replacement 
\begin{equation}
\big[ j_{\vec k}^Q, A_{{\vec q}-{\vec k}} \big] \Rightarrow 
\big[ H, A_{\vec q} \big], \ \ \ \ \ 
\big[ j_{\vec k}^Q, A_{\vec q}^\dag \big] \Rightarrow 
\big[ H, A_{{\vec q}-{\vec k}}^\dag \big], 
\end{equation}
in the limit of vanishing external momentum, ${\vec k}\rightarrow 0$. 

Irrespective of the range of the interaction 
we obtain the Ward identity 
\begin{equation}
\sum_{\mu=0}^3 k_\mu M_\mu^Q(q+k,q) = 
(q_0+k_0) D(q+k) - q_0 D(q), \label{WI-Q}
\end{equation}
for any pairs. 
This result is consistent with the Ward identity for phonons~\cite{Ono}. 
Here $D(q)$ is the Fourier transform of the fluctuation propagator 
\begin{equation}
D(x,y) = 
- i \big\langle T \{ A(x) A^\dag(y) \} \big\rangle, 
\end{equation}
for particle-particle or particle-hole pairs. 
The pair with four-momentum $q$ carries the energy of $q_0$. 

\section{Conclusion}

In the discussion of the Ward identity 
the commutation relation plays the central role. 

In the case of the electric current vertex 
it picks up the charge of the pair. 
Thus it is concluded that charge-neutral pairs do not couple 
to electromagnetic field. 
Namely charge- and spin-density fluctuations do not carry charge. 
On the other hand, 
Cooper pairs carrying charge $2e$ 
couple to electromagnetic field. 

In the case of the heat current vertex 
it picks up the energy of the pair. 


\end{document}